# The Three Books of Science

Peter Sheridan Dodds

Computational Story lab | Vermont Complex Systems Institute
Department of Computer Science | Vermont Advanced Computing Center
University of Vermont, Burlington, VT 05405, USA

Santa Fe Institute, Santa Fe, NM 87501, USA

October 31, 2025

**Abstract**

We venture that the long evolution of science may be viewed as unfolding over three blurred epochs. The first epoch spans the slow, haphazard, error-ridden realization of scientific truths along with foundational scientific methods. The second epoch covers the discovery that everything is formed from atoms and other building blocks. And in the age of computation, the third epoch affords the science of complex systems.

**Spoiler**

The science of complex systems has always been where science was headed. It's the third and final book in the trilogy that is science. We're talking about basic, fundamental science, on par with cosmology and quantum mechanics.

**The first book**

We took thousands of years to develop science in the first place. Physics has become enormously successful but that was after thousands of years of measurement and centuries upon centuries of being harmoniously wrong about the basics. Circles instead of ellipses.

Apart from the "Science is Hard" issue, religion made things a little tricky along the way. When it really mattered for Team Belief, the Number Boffins and religious leaders could work together. The adjustment from the Julian Calendar to the Gregorian is one of the great examples. Measurement and continued fractions—pure joy—brought us to the calendar we use now where every 4 years we have a leap year, every 100 we do not, but every 400 we do—the exception to the exception to the exception.[1] The year 2000 was gigantic for The CalendArmy[2] but went unnoticed by the calendrically indifferent—almost everyone—who were largely worried that their digital clocks might explode.

This first book is a big mess. Much of it was spoken before the invention of books and papers. Most of the pages are lost. If we did have all the books, most pages would be scribbled out or covered in warnings. These unwritten and lost pages of the first book would all be filled with madnesses.

"Today, we covered Og in feathers and threw him off the cliff. Og will be okay. Probably. We need more feathers. And honey"

"We covered Grog in two times feathers and threw him off the cliff. More feathers."

"We thew Splog off the cliff with no feathers. Not an experiment. Splog

---

[1]Except for the Possibly Accidental Calendrical Separatists on the island of Foula for what seems to have been a tiny mistake in 1900.
[2]Every true fandom has to have a name.

wouldn't shut up about how hunting is better than flying. He'll be okay. Unfortunately."

Several pages later: "Maybe we need a bigger cliff."

Later: "We found a cliff above water."

Even later: "We found a cliff that's not that far above water."

"The birds are too fast, we can't catch them."

"Splog is our best hunter and he's still limping," said Phrog, reproachfully.

"All right, back to the drawing cave. Let's try breathing underwater again. Who wants to go first?"

The survivor bias for knowledge. But all of this is all in retrospect only—knowledge is hard gained.

We discovered many basic things over and over before finally starting to build a platform of knowledge. The right angle triangle with squares business is completely misnamed as Pythagoras's theorem. Stigler's law of eponymy states that we routinely wrongly name scientifc laws, mis-attributing them by accident, lack of knowledge, or because someone wanted to be famous [1]. Always with the fraud. Of course, Stigler's law of eponymy is itself misnamed and should be Robert Merton's law of eponymy. Stigler misnamed his own law to add further evidence, which is all very cheeky and naughty. Reality is reality but the mapping of reality, however good eventually, is socially constructed [2].

So in the last few hundred years, we've started to pull things together by formally sharing knowledge through papers and journals (but let's not discuss peer review right now). And in the last 30 years, our ability to sort through scientific knowledge has flipped from small-scale to large-scale. We now have a different hardly search problem and it can seem impossible but we've made more connections across fields. At the heart of this has been data, and its the move from data scarce to data rich that's helped us find which patterns and mechanisms transcend fields.

**The second book**

Once science started to become what we would now recognize as science—the word 'scientist' was only coined in 1830 or so because of consilience and the -ist is from artist—we took several hundred years to figure out the existence of atoms, DNA, viruses, the human genome, and that systems are built out of or coded by small indivisible units. All of this is still new—it's only in the early 1900s that physicists finally agreed that atoms exist. Most of them. In the year 1900, atoms were viewed as an idiotic conception by many. There's a long history of the anti-atomicism. Around 2400 years earlier, Democritus, one of those Ancient Greek Thinker Boffins, had come up with the concept of the atom: a – temnein (not cut).[3] That everything is made up of indivisible units. Plato apparently wanted to burn all of Democritus's books—A Distributed Denial of Science attack—which could all be just a good story. While his brand remains strong, Plato got a few things wrong. Chemists brought about the what we might call the Golden Age of Reductionism, which would start around 1700 and last for hundreds of years. With some completely reasonable excursions into alchemy.[4]

In 1905, Einstein had a Pretty Good Year (PGY) and one of his papers showed that an atomic model of little things bouncing off other little things could give rise to Brownian motion [3]. In 1908, Jean Perrin backed up Einstein's theory with experiments and the atom part of the second book was starting to take shape [4]. A moment for Boltzmann. In 1911, which some refer to as 1BT,[5] Rutherford's experiments showed that J.J. Thompson's plum pudding model of the atom was not at all correct, or at least that the pudding was prepared with some difficulty and would be rather unappetizing.[6,7]

---

[3]Not recorded: Splog also thought things were made of little things but he was socially constrained to behaving like a hunter.

[4]Absolutely fair play. We had transmuted all kinds of stuff beforehand into something people highly valued. Alcohol. We can make meringues out of water from a can of chickpeas.

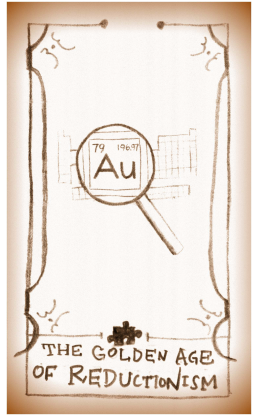

Figure 1: The Reductioning. There Can Be Only Some … Parts.

We'll keep writing the second book as long as there are so many more building blocks to find. It's a long list, one of those heavy tail things. The rare are legion. All of the molecules, all of the species of everything on Earth. But because evolutionary processes (which includes engineers and mad scientists) will keep creating new building blocks, we'll have to keep updating the second book. New viruses are always fun. LEGOS were only discovered in 1949 by some enterprising Plastics Boffins in Denmark. And we keep finding things are made out of LEGOS. And increasingly we've discovering it's not just cake that's made out of cake. This is a separate discussion, but humans, and, by inclusion, scientists—well, mostly—are really, really into food.

---

[5] A year before the Fight of the Century, The Commotion in the Ocean, The Get-Frantic in the At-lantic, The Metaphor far out from the Seashore: Iceberg versus The Titanic.

[6] To this day, no one has ever made a successful model-of-the-atom showstopper on a national baking show.

[7] Rule #37: Don't do science on an empty stomach.

Rutherford—in a solid instance of a New Zealander going lower than Australians in their behavior—allegedly said "All science is either physics or stamp collecting." But the second book is not just a philatelical festival. It is absolutely fundamental, bedrock science, full of wonder and magic. The atoms of course weren't the end for physicists who have very cleverly kept the money rolling in and continued to give writers material for preposterous but great science fiction (see, for example, sophons). The Tiny Things Boffins first came up with subatomic particles, then strings, then dark matter, and dark energy. They may well have played us for absolute fools, but we now have the word quark—it can't be the darkest timeline.

Here's Super Boffin Feynman in Volume 1 [5]:

"If, in some cataclysm, all of scientific knowledge were to be destroyed, and only one sentence passed on to the next generation of creatures, what statement would contain the most information in the fewest words?

"I believe it is the atomic hypothesis that all things are made of atoms—little particles that move around in perpetual motion, attracting each other when they are a little distance apart, but repelling upon being squeezed into one another.

"In that one sentence, you will see, there is an enormous amount of information about the world, if just a little imagination and thinking are applied."

**The third book is about complex systems—how things fit together, how things stay together, how things fall apart**

What's a complex system? A complex system is a distributed system of many interconnected parts, not governed by any dominant centralized control, possibly with networked structures, and, crucially, one that has emergent behavior [6, 7]. Emergent phenomena are everywhere, from raw physical systems up through all levels of increasingly algorithmic and computational phenomena.[8]

Emergence is the key: We see patterns and structures and dynamics at the system scale that are not present in the parts.

- There's no hurricane in a water molecule.
- No financial collapse in a dollar bill.
- No justice in a carbon atom.
- No Shakespeare play in 26 letters and some punctuation.

Good luck searching for Romeo and Juliet in Borges's fictional hexagons-are-the-bestagons-based library. Disappointingly, you will continue to find something like Robeo and Muliet in the functional infinity of adjacent texts, if you can ever make it that close.

Emergence is so confounding that ideas like a homunculus have found their way into boffin brains, which is some kind of inception of self.

**A modest manifesto**[9]

Complex systems are everywhere. The systems that matter most are undisciplinary, increasingly of global scale, and involve the behavior of people, physics, and algorithms. We have social systems, information systems, economies, energy systems, food systems, pandemics, political systems, entertainment, sports, artificial intelligence systems, and of course, real stupidity systems [8].

---

[8]The advent of life is the advent of algorithms.
[9]Remanifested from many manifestations of Principles of Complex Systems I and II, https://pdodds.w3.uvm.edu/pocsverse/.

We are in a renaissance time in science where understanding complex systems is now available to us, for three reasons:

**One:** There are universal mechanisms, patterns, and dynamics that appear across disparate systems. Knowledge of systems is not bespoke and siloed. It could have been! That there is one set of equations that describes the dynamics of water, air, and the Earth, and maple syrup was not obvious (but nothing is obvious).

**Two:** Our ability to measure everything continues to expand enormously, and we continue to move fields from being data scarce to data rich, from non-computational to computational. Computational physics, computational biology, computational medicine, computational social sciences, computational humanities.[10]

**And three:** We can computationally simulate real complex systems and experiment with hypothetical ones, all with increasing power. Advanced computing facilities are our blackboards.

And we're finally seeing this transition in the big awards: The 2024 Nobel prizes in physics and chemistry went to "computational scientists" whose work[11] created complex systems—neural networks and AI—or simulated the once impossible like the folding of proteins. Not everyone was happy about this of course, but you know, scientists.[12]

---

[10] A field being data rich does not mean the end of small data work. Computational sciences are complementary to not obliterating of a field's work. That said, most no-data or small-data theories will not survive contact with data. Which is to be expected. Kudos again to Tiny Things Boffins.

[11] And the work of many, many others

[12] Non-starter alternate plan: Honor what has been uncovered by science and not the discoverers. Best new supporting species.

We will always need to study the specific and the broad. Viruses, individual species, the one offs, the irreproducible accidents.

But to write the third book of science, we know that we need a postdisciplinary, computational science of complex systems. We will never finish the second and third books, but their writing is the inherent mandate of science, and one we must navigate with a true moral compass.[13]

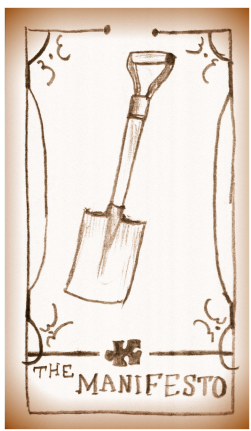

Figure 2: Grab a shovel. Lots to do.

---

[13]For the Greater Good but not for the "For the Greater Good" kind of Greater Good.

# References

[1] S. M. Stigler. Stigler's law of eponymy. *Transactions of the New York Academy of Sciences*, 39:147–157, 1980.

[2] P. L. Berger and T. Luckmann. *The social construction of reality: A treatise in the sociology of knowledge*. Penguin UK, 1991.

[3] A. Einstein. Über die von der molekularkinetischen theorie der wärme geforderte bewegung von in ruhenden flüssigkeiten suspendierten teilchen. *Annalen der Physik*, 322:549–560, 1905.

[4] J. B. Perrin. Mouvement brownien et réalité moléculaire. *Annales de Chimie et de Physique*, 18(8):1–114, 1909. Translated into English by F. Soddy as *Brownian Movement and Molecular Reality*.

[5] R. P. Feynman, R. B. Leighton, and M. Sands. *The Feynman Lectures on Physics, Volume I*. Addison–Wesley, Reading, MA, 1963.

[6] G. H. Lewes. *Problems of Life and Mind, Vol. II: The Physical Basis of Mind*. Trübner and Co., London, 1875. See Chapter V, "The Nature of Emergent Effects".

[7] P. W. Anderson. More is different. *Science*, 177(4047):393–396, 1972.

[8] C. M. Cipolla. The basic laws of human stupidity. In *Allegro ma non troppo. Saggi satirici*. Società editrice il Mulino, Bologna, Italy, 1988. Originally written in English in 1976 for private circulation. "Le leggi fondamentali della stupidità umana," included in this volume.